\documentclass[pra, showpacs, twocolumn, a4paper, nofootinbib]{revtex4}
\usepackage{graphicx}
\usepackage{amsmath, amsfonts, amssymb, bm}
\usepackage[]{psfrag}
\psfragscanoff 

\usepackage{verbatim}
\usepackage{bm}

\begin{document}

\newcommand{\ket}[1]{| #1 \rangle}
\newcommand{\bra}[1]{\langle #1 |}

\title{Temporal and diffraction effects in entanglement creation in an optical cavity}
\author{Sonny \surname{Natali}}
\author{Z. \surname{Ficek}}
\email{ficek@physics.uq.edu.au}
\affiliation{Department of Physics, School of Physical Sciences, \\
The University of Queensland, Brisbane, Australia 4072}

\date{\today}

\begin{abstract}
A practical scheme for entanglement creation between distant atoms located inside a single-mode optical cavity is discussed.
We show that the degree of entanglement and the time it takes for the entanglement to reach its optimum value is a sensitive function 
the initial conditions and the position of the atoms inside the cavity mode. It is found that the entangled properties of the two atoms 
can readily be extracted from dynamics of a simple two-level system. Effectively, we engineer two coupled qubits whose the dynamics are 
analogous to that of a driven single two-level system. It is found that spatial 
variations of the coupling constants actually help to create transient entanglement which may appear on the time scale much longer than 
that predicted for the case of equal coupling constants. When the atoms are initially prepared in an entangled state, they may remain 
entangled for all times. We also find that the entanglement exhibits an interesting phenomenon of diffraction when the the atoms are 
located between the nodes and antinodes of the cavity mode. The diffraction pattern of the entanglement varies with time and we explain 
this effect in terms of the quantum property of complementarity, which is manifested as a tradeoff between the knowledge of energy of 
the exchanged photon versus the evolution time of the system. 
\end{abstract}


\maketitle

\section{Introduction}

The study of practical schemes for creation of quantum entanglement between atoms (or ions) is the most active area in quantum optics 
and quantum information science~\cite{nil}. Different schemes have been proposed including atom trapped inside a single mode 
cavity~\cite{cz94,tur,wang}, or inside two separate cavities~\cite{dc00,zm04,adr04,mb04,ye04,kc04,ch05,smb06}. One of the most popular 
scheme involves two two-level atoms located (trapped) within a single-mode cavity field. It has been demonstrated that the entanglement 
could in principle be created through a continuous observation of the cavity field~\cite{phbk} or through dispersive atom-cavity field 
interactions~\cite{gerry,zg00,xl05}, thereby creating a strong "action at a distance". The approach used is straightforward: Provided 
no photon is leaking through the cavity mirrors or no photon is exchanged between the atoms 
and the cavity field, a pure entangled state between the two atoms results. However, all of these procedures for generating entangled 
atoms have suffered from a common handicap: their choice of equal coupling strengths of the atoms to the cavity mode. The difficulty
is that the entanglement depends on the the coupling constant between the atoms and cavity mode which depends, in turn, on the 
location of the atoms in the cavity mode. In a standing wave cavity, one can achieve the equal coupling constants by locating the 
atoms precisely at the antinodes of the cavity field, or by sending slowed (cooled) atoms through a cavity antinode in the direction 
perpendicular to the cavity axis. This is a relatively easy task at microwave frequencies and, in fact, detection 
of entangled atoms have already been performed on a beam of Rydberg atoms traversing a superconducting microwave cavity~\cite{os01}. 
However, at optical frequencies this task may well be hard to achieve. In practice, the dipole coupling constants vary with the 
location of the atoms in the cavity mode. For example, in a standing-wave structure of the cavity mode, the coupling constant varies 
with the position of the atom as~\cite{yariv}
\begin{eqnarray}
      g(\vec{r})\equiv g(r,z) = g_{0}{\rm e}^{-r^{2}/w_{0}^{2}}\cos\left(kz\right) , \label{1.1} 
\end{eqnarray}
where $z$ determines location of the atom along the cavity axis, $k=2\pi/\lambda$ is the wave number of the field,
$r= (x^{2}+y^{2})^{1/2}$ is the distance of the atom from the cavity axis, and $w_{0}$ is the mode waist. The coupling constant 
reaches maximum value $g_{0}$ when the atom is located on the cavity axis at an antinode of the standing wave. In practice one would 
like the varying coupling constant $g(\vec{r})$ to coincide with $g_{0}$. However, because of the small wavelength, locating the atoms 
precisely at the antinodes of the cavity mode and thereby eliminating the variation of $g(\vec{r})$ with position is very difficult 
in an optical cavity. This may change the physics completely and thus suggests that the problem of creating entanglement between distant 
atoms in an optical cavity is a significant experimental challenge.

In this paper, we are concerned with the optical frequency regime and investigate the mechanism involved in creation 
of entanglement between distant atoms coupled to a single-mode cavity field. Our treatment closely follows the approach that was used
Refs~\cite{gerry,zg00,xl05}, but with one essential difference. We include a possible variation of the coupling constant $g(\vec{r})$ 
with the location of the atoms in a standing-wave cavity mode. We are particularly interested in the consequences 
of this variation on entanglement creation between the atoms, since this will be very pertinent to any practical experimental 
arrangements as the distances involved are very small. First of all, we derive the general master equation for the reduced 
density operator $\rho_{s}$ of two two-level atoms coupled to a single mode cavity field.
Our approach holds for atoms at rest or slowly moving through a single-mode cavity. For sufficiently cold or slow 
atoms, radiative equilibrium is reached with an essentially fixed coupling constant, $g(\vec{r})$, at every point inside the cavity, 
permitting studies of entangled properties of the system without performing the average over random locations and atomic 
dipole orientations. We assume the atoms are far enough apart that the direct dipole-dipole coupling or other direct interactions
between the atoms can be neglected. 
This also allows a selective preparation of the atoms such that a given set of initial conditions for the atomic states is achieved.
We solve the master equation for two atoms coupled to a cavity mode in the limit of a large detuning of the cavity mode 
from the atomic resonance frequency. This enables to eliminate the cavity mode and obtain a new master equation in which we will 
recognize some terms equivalent to the dipole-dipole interaction between the atoms and to the Stark shift of the atomic resonance 
frequencies. The analogy of this system with that of a single two-level atom driven by a detuned coherent field is exploited and 
discussed. This analogy provides a simple description of the process of entanglement creation and leads to a useful pictorial 
representation of the system in terms of the Bloch vector model.
To quantify the degree of entanglement, we use the concurrence that is the widely accepted measure of two-atom entanglement. Simple 
analytical expressions are obtained for the concurrence that are valid for arbitrary initial conditions and arbitrary positions of 
the atoms inside the cavity mode. We obtain the interesting result that spatial variations of the coupling constants actually help 
to create transient entanglement which may appear on the time scale much longer than that predicted for the case of equal coupling 
constants. We explain this effect in terms of the degree of localization of the energy induced in the field by the interacting atoms.
Moreover, we find that for an imperfect location of the atoms inside the cavity mode the entanglement exhibits an interesting 
time-dependent diffraction phenomenon.

\section{Master equation}

The system we consider consists of two identical two-level atoms (qubits) with upper levels $\ket{e_{i}}$,\ $(i=1,2)$, lower 
levels $\ket{g_{i}}$, and separated by energy $\hbar\omega_{0}$. The atoms are coupled to a standing-wave cavity mode with the 
position dependent coupling constants $g(\vec{r}_{i})$, and damped at the rate $\gamma$ by spontaneous emission to modes 
other than the privileged cavity mode.
The cavity mode is damped with the rate $\kappa$ and its frequency $\omega_{c}$ is significantly detuned from the atomic transition 
frequency $\omega_{0}$, so there is no direct exchange of photons between the atoms and the cavity mode. The behavior of the system 
is described by the density operator $\rho$,  which in the interaction picture satisfies the master equation
\begin{eqnarray}
\frac{\partial \rho}{\partial t} = -\frac{i}{\hbar}[H,\rho]  
+\frac{1}{2}\gamma {\cal L}_{a}\rho +\frac{1}{2}\kappa {\cal L}_{c}\rho ,\label{2.1} 
\end{eqnarray}
where
\begin{eqnarray}
     H = \hbar\sum_{j=1}^{2}\left[g(\vec{r}_{j})a S_{j}^{+}{\rm e}^{-i\Delta t} + {\rm H.c.}\right] \label{2.2}  
\end{eqnarray}
describes the interaction between the cavity field and the atoms,
\begin{eqnarray}
     {\cal L}_{c}\rho = \left(2a\rho a^{\dagger} - a^{\dagger}a\rho - \rho a^{\dagger}a\right) ,\label{2.3}
\end{eqnarray}
and
\begin{eqnarray}
  {\cal L}_{a}\rho = \sum_{j=1}^2\left(2S_{j}^{-}\rho S_{j}^{+} 
  - S_{j}^{+}S_{j}^{-}\rho - \rho S_{j}^{+}S_{j}^{-}\right)  .\label{2.4}
\end{eqnarray}
are operators representing the damping of the atoms by spontaneous emission and of the cavity field by the cavity decay, respectively.
The operators $S_{j}^{+}$  and $S_{j}^{-}$ are the raising and lowering operators of the $j$th atom, $S_{j}^{z}$ describes its energy, 
$a$ and $a^{\dagger}$ are the cavity-mode annihilation and creation operators, $\Delta =\omega_{c}-\omega_{0}$ is the detuning of 
the cavity-mode frequency from the atomic transition frequency, and $\vec{r}_{j}$ is the position coordinate of the $j$th atom 
within the cavity mode.

The atoms located at different positions may experience different coupling constants, that is $g(\vec{r}_{1})\neq g(\vec{r}_{2})$.
Because it is precisely the effect of unequal coupling constants that interest us most here, we choose the reference frame such that 
\begin{eqnarray}
g(\vec{r}_{1}) = g_{0} ,\quad {\rm and} \quad g(\vec{r}_{2})= g_{0}\cos\left(kr_{12}\right) , \label{2.5} 
\end{eqnarray}
where $r_{12} =z_{2}-z_{1}$ is the distance between the atoms. This choice of the reference frame corresponds to a situation 
where atom 1 is kept exactly at an antinode of the standing wave and the atom 2 is moved through successive nodes and antinodes of 
the standing wave. This choice, of course, involves no loss of generality.

We also assume that the atoms are stationary during the interaction with the cavity mode, i.e. the distance between the atoms is 
independent of time (the Raman-Nath approximation). This is a good approximation for many experiments on cooling of trapped atoms, 
where the storage time of the trapped atoms is long, so that they are essentially motionless and lie at known and controllable 
distances from one another~\cite{wine}. 

In order to study the dynamics of the system, we introduce density-matrix elements with respect to the cavity field mode, 
denoting $\bra n \rho \ket m$ by $\rho_{nm}$, and find from the master equation~(\ref{2.1}) that the populations of the two lowest 
energy levels and coherence between them satisfy the following equations of motion
\begin{eqnarray}
  \dot{\rho}_{00} &=& -i\sum_{j=1}^{2}g_{j}\left(S_{j}^{+}\tilde{\rho}_{10} - \tilde{\rho}_{01}S_{j}^{-}\right) 
  +\frac{1}{2}\gamma {\cal L}_{a}\rho_{00} +\kappa\rho_{11} ,\nonumber \\
  \dot{\tilde{\rho}}_{01} &=& i\Delta\tilde{\rho}_{01} - i\sum_{j=1}^{2} g_{j}\left(S_{j}^{+}\rho_{11} 
  - \rho_{00}S_{j}^{+}\right) \nonumber \\
  && +\frac{1}{2}\gamma {\cal L}_{a}\tilde{\rho}_{01} -\frac{1}{2}\kappa\tilde{\rho}_{01} ,\nonumber \\
  \dot{\tilde{\rho}}_{10} &=& -i\Delta\tilde{\rho}_{10} 
  + i\sum_{j=1}^{2} g_{j}\left( \rho_{11}S_{j}^{-} - S_{j}^{-}\rho_{00}\right) \nonumber \\
  && +\frac{1}{2}\gamma {\cal L}_{a}\tilde{\rho}_{10} -\frac{1}{2}\kappa\tilde{\rho}_{10} ,\nonumber \\
  \dot{\rho}_{11} &=& -i\sum_{j=1}^{2}g_{j}\left(S_{j}^{-}\tilde{\rho}_{01} - \tilde{\rho}_{10} S_{j}^{+}\right) \nonumber \\
  && +\frac{1}{2}\gamma  {\cal L}_{a}\rho_{11} -\kappa\rho_{11} ,\label{2.6}
\end{eqnarray}
where $g_{j}\equiv g(\vec{r}_{j})$ and $\tilde{\rho}_{nm}$ are slowly varying parts of the coherences, 
$\tilde{\rho}_{01} = \rho_{01}\exp(i\Delta t)$ and $\tilde{\rho}_{10} = \rho_{10}\exp(-i\Delta t)$. We have considered only the 
two lowest energy levels as in the limit of a large detuning $\Delta$ only the ground state $(n=0)$ and
the one-photon state $(n=1)$ of the cavity mode are populated.

Now, we explicitly apply the adiabatic approximation that for a large detuning, the coherences $\tilde{\rho}_{01}$ and 
$\tilde{\rho}_{10}$ vary slowly in time, so we can assume that $\dot{\tilde{\rho}}_{01}\approx 0$ and 
$\dot{\tilde{\rho}}_{10}\approx 0$. In this case, we find from (\ref{2.6}) that in the limit of $\Delta \gg g_{j}\gg\gamma ,\kappa$
\begin{eqnarray}
  \tilde{\rho}_{01} &\approx& \frac{1}{\Delta}\sum_{j=1}^{2}g_{j}\left(S_{j}^{+}\rho_{11} - \rho_{00}S_{j}^{+}\right) 
  ,\nonumber \\
  \tilde{\rho}_{10} &\approx& \frac{1}{\Delta}\sum_{j=1}^{2} g_{j}\left( \rho_{11}S_{j}^{-} - S_{j}^{-}\rho_{00}\right) .\label{2.7}
\end{eqnarray}
Knowledge of the coherences $\tilde{\rho}_{01}$ and $\tilde{\rho}_{10}$ allows us to derive the master equation for the reduced 
density operator of the atoms. It is done in the following way. First we substitute (\ref{2.7}) into (\ref{2.6}), and after
neglecting the population $\rho_{11}$, as the cavity mode will never be populated, we find
\begin{eqnarray}
 \dot{\rho}_{00} = \frac{i}{\Delta}\sum_{i,j=1}^{2}g_{i}g_{j}\left[S_{i}^{+}S_{j}^{-},\rho_{00}\right] 
 +\frac{1}{2}\gamma {\cal L}_{a}\rho_{00} .\label{2.8}
\end{eqnarray}
Since $\rho_{00} = {\rm Tr}_{F}(\rho) = \rho_{s}$ is the reduced density operator of the atoms, we obtain the master equation 
for the density operator of the atoms
\begin{eqnarray}
  \frac{d\rho_{s}}{dt} &=& i\sum_{i=1}^{2}\delta_{i}\left[S_{i}^{+}S_{i}^{-},\rho_{s}\right] 
    + i\sum_{i\neq j=1}^{2}\Omega_{ij} \left[S_{i}^{+}S_{j}^{-},\rho_{s}\right] \nonumber \\ 
  && + \frac{1}{2}\gamma {\cal L}_{a}\rho_{s} ,\label{2.10}
\end{eqnarray}
where
\begin{eqnarray}
  \delta_{i} = \frac{g_{i}^{2}}{\Delta} ,\quad {\rm and}\quad \Omega_{ij} = 
  \Omega_{ji} = \frac{g_{i}g_{j}}{\Delta} .\label{2.11}
\end{eqnarray}
The first two terms in the master equation (\ref{2.10}) depend on the position coordinate of the atoms and give rise to frequency 
shifts of the atomic levels and the coupling between the atoms, respectively. The third term represents the damping of the atoms through 
the interaction with the environment.
The parameter $\delta_{i}$ represents the shift in energy separation of the levels of the $i$th atom due to the dispersive 
interaction with the cavity mode. It is an analog of a dynamic Stark shift. One can easily see from the structure of the first term in 
the master equation that 
\begin{eqnarray}
  && i\sum_{i=1}^{2}\delta_{i}\left[S_{i}^{+}S_{i}^{-},\rho_{s}\right] 
   = i\sum_{i=1}^{2}\delta_{i}\left[S_{i}^{z} + \frac{1}{2},\rho_{s}\right] \nonumber \\
  && = i\sum_{i=1}^{2}\delta_{i}\left[S_{i}^{z},\rho_{s}\right] ,\label{2.12}
\end{eqnarray}
which clearly shows that this term is an analog of the energy shift term. Thus in the interaction picture used here, it represents 
a shift of the atomic energy levels. We note that due to nonequivalent positions of the atoms, the shift of the energy levels is 
different for different atoms. 

The multi-atom term $\Omega_{ij}$ represents the shift in energy separation of the levels of atom $i$ due to its interaction with
the atom $j$ through the cavity mode. If the atoms are located at antinodes of the standing wave, the term $\Omega_{ij}$ is maximal,
whereas $\Omega_{ij}=0$ if at least one of the atoms is located at a node of the standing wave. From the structure of the second 
term in Eq.~(\ref{2.10}) one can recognize that $\Omega_{ij}$ is an analog of the familiar dipole-dipole interaction between 
the atoms~\cite{dic,ft02}. This shows that the interaction of the atoms with a detuned cavity fields produces a structure in the master 
equation analogous to the dipole-dipole interaction between the atoms.

The above procedure shows that the adiabatic elimination of the cavity mode creates a shift of the atomic transition frequencies 
and an effective interaction between two distant atoms. Thus, the dynamics of the system composed of two identical atoms in 
nonequivalent positions in the cavity mode is equivalent to those of two non-identical atoms of different transition frequencies.
In other words, the procedure is an example of how one can ``engineer'' the dipole-dipole 
interaction between distant atoms. It is now easy to understand why two independent atoms coupled to a strongly detuned cavity mode
can exhibit entanglement. Simply, the reduced system is equivalent to two atoms coupled through the induced dipole-dipole interaction. 
Exactly, this process gives rise to the entanglement.

We point out in passing that despite of the presence of the coherent dipole-dipole interaction term, the master equation (\ref{2.10}) 
is not fully equivalent to the master equation of two collective interacting atoms~\cite{fs05}. This is because there is no contribution
from the cross-damping terms involving dipole operators of two different atoms. In other words, the interaction with a strongly detuned 
cavity mode does not create the collective damping of the atoms. As a result, the atoms interact independently 
with the environment, so that the system does not evolve to a dark state characteristic of the completely collective 
system~\cite{ft02,afs,fs05}. The cross-damping terms would appear in the master equation if one assumes the near resonant interaction,
$\Delta \approx 0$, and the "bad-cavity" limit of $\kappa \gg g_{j}\gg \Delta$~\cite{wang,gm97}.

\section{Equivalent two-level dynamics}

The question we are interested in concerns the consequences of the spatial variation of the coupling constant $g$ on the 
entanglement creation between two atoms located in a strongly detuned single-mode cavity field. To answer this question we consider
the evolution of the diagonal density matrix elements which correspond to the occupation probabilities of the energy levels of the 
two-atom system. Using the master equation (\ref{2.10}), we find the following equations of motion 
\begin{eqnarray}
  \dot{\rho}_{11} &=& \gamma - \gamma \left(\rho_{11} + \rho_{44}\right) , \label{3.1}\\
  \dot{\rho}_{44} &=& -2\gamma\rho_{44} ,\label{3.2} \\
  \dot{\rho}_{22} &=& -\gamma\rho_{22} + \gamma\rho_{44} 
  + i\Omega_{12}\left(\rho_{23} - \rho_{32}\right) , \label{3.3}\\
  \dot{\rho}_{33} &=& -\gamma\rho_{33} + \gamma\rho_{44} 
  - i\Omega_{12}\left(\rho_{23} - \rho_{32}\right) ,\label{3.4} 
\end{eqnarray} 
while the off-diagonal density matrix elements that are coupled to the diagonal elements obey the equations
\begin{eqnarray}  
  \dot{\rho}_{23} &=& -\left(\gamma -i\delta_{12}\right)\rho_{23} 
  + i\Omega_{12}\left(\rho_{22} - \rho_{33}\right) , \label{3.5}\\
  \dot{\rho}_{32} &=& -\left(\gamma +i\delta_{12}\right)\rho_{32} 
  - i\Omega_{12}\left(\rho_{22} - \rho_{33}\right) , \label{3.6}
\end{eqnarray}
where we have used the standard direct-product basis given by
\begin{eqnarray}
 \ket{1} &=& \ket{g_{1}}\ket{g_{2}} ,\quad \ket{2} = \ket{g_{1}}\ket{e_{2}} ,\nonumber \\
 \ket{3} &=& \ket{e_{1}}\ket{g_{2}} ,\quad \ket{4} = \ket{e_{1}}\ket{e_{2}} .\label{3.7}
\end{eqnarray}
Here, $\delta_{12} =\delta_{1}-\delta_{2}$ is a difference between the single-atom Stark shifts. This parameter is of central 
importance here as it determines the relative variation of atomic transition frequencies with position of the atoms inside the 
cavity mode. In the special case of $g_{1}=g_{2}$, the parameter $\delta_{12} =0$, but this can happen only when the atoms are 
in equivalent positions inside the mode.

It is easy to see that the set of the coupled equations~(\ref{3.1})--(\ref{3.6}) can be split into two independent sets of 
equations of motion, which can be written in the form
\begin{eqnarray}
  \dot{\rho}_{11} &=& \gamma - \gamma \left(\rho_{11} + \rho_{44}\right) , \nonumber \\
  \dot{\rho}_{44} &=& -2\gamma\rho_{44} , \nonumber \\
  \dot{\rho}_{++} &=& -\gamma\rho_{++} +2\gamma\rho_{44} ,\label{3.8} 
\end{eqnarray}
and
\begin{eqnarray}
  \dot{u} &=& -\gamma u + \delta_{12}v ,\nonumber \\
  \dot{v} &=& -\gamma v - \delta_{12} u  -2\Omega_{12} w ,\nonumber \\
  \dot{w} &=& -\gamma w + 2\Omega_{12}v ,\label{3.9}
\end{eqnarray}
where $\rho_{++}=\rho_{22}+\rho_{33} $, $u= \rho_{23}+\rho_{32}$, $v= i(\rho_{23}-\rho_{32})$, and $w= \rho_{22}-\rho_{33}$. 
Note that the equations of motion (\ref{3.9}) 
are the exact equivalent of the optical Bloch equations of a two-level system driven by a detuned coherent field, where the 
dipole-dipole interaction $\Omega_{12}$ couples to the levels like a coherent field, and $\delta_{12}$ appears as a detuning of 
the field from the driven transition. The upper and lower energy levels $\ket 2$ and $\ket 3$ thus show a dynamics that is analogous 
to that of a driven two-level system. It should be pointed out that the analogy is not absolute because, in contrast to the case of 
a single two-level atom driven by a detuned laser field, $\Omega_{12}$ and $\delta_{12}$ in Eq.~(\ref{3.9}) depend on the same 
parameters $g_{1}$ and $g_{2}$, i.e. on the coupling constants of the atoms to the cavity field. Consequently, the parameters
$\Omega_{12}$ and $\delta_{12}$ are not independent.

The dynamics of the effective two-level system can be easily visualized in the Bloch vector model~\cite{al75}. In this model, the system 
and the driving field are represented by vectors in a three-dimensional space, and the time evolution is simply 
visualized as a precession of the system-state vector about the driving field. In terms of the Bloch vector, Eq.~(\ref{3.9}) can be 
written as
\begin{eqnarray}
   \frac{d\vec{B}}{dt} = -\gamma \vec{B} +\vec{\Omega}_{B}\times \vec{B}  ,\label{3.10}
\end{eqnarray}
where $\vec{\Omega}_{B}= (-2\Omega_{12},0,\delta_{12})$ is the pseudofield vector and $\vec{B}= (u,v,w)$ is the Bloch vector.
The quantities $u$ and $v$ are, respectively, the real and imaginary parts of the coherence between the levels $\ket2$ and $\ket 3$, 
and $w$ is the population inversion. In the present problem, the coherence is induced by the dipole-dipole interaction
$\Omega_{12}$ which, as we have already pointed out, plays a role similar to the Rabi frequency of the coherent interaction between the 
levels, i.e. represents a rate at which one quantum of excitation is exchanged between the atoms. It should be noted that there is no 
electric-dipole moment between the levels $\ket 2$ and $\ket 3$, so there are no radiative transitions between the levels of the two-level 
system. The damping rate $\gamma$ that appears in Eq.~(\ref{3.9}) represents spontaneous decay out of the system to the auxiliary 
level~$\ket 1$.
 
Since the $u$ and $v$ components of the coherence are related to the interaction between the atoms, their properties should be reflected 
in the entanglement between the atoms.
In order to determine the amount of entanglement between the atoms and the entanglement dynamics, we use concurrence that 
is the widely accepted measure of entanglement. The concurrence introduced by Wootters~\cite{woo} is defined as
\begin{eqnarray}
  {\cal C} = \max\left(0,\sqrt{\lambda_{1}}-\sqrt{\lambda_{2}}-\sqrt{\lambda_{3}} -\sqrt{\lambda_{4}}\,\right) ,\label{3.11}
\end{eqnarray}
where $\{\lambda_{i}\}$ are the the eigenvalues of the matrix
\begin{equation}
  R=\rho_{s}\tilde{\rho}_{s} ,\label{3.12}
\end{equation}
with $\tilde{\rho}_{s}$ given by
\begin{equation}
  \tilde{\rho}_{s} = \sigma_{y}\otimes\sigma_{y}\,\rho^{*}_{s}\,\sigma_{y}\otimes\sigma_{y} ,\label{3.13}
\end{equation}
and $\sigma_{y}$ is the Pauli matrix. The range of concurrence is from 0 to 1. For unentangled atoms ${\cal C}=0$ whereas ${\cal C}=1$ 
for the maximally entangled atoms. In terms of the density matrix elements, the concurrence is given by
\begin{equation}
  {\cal C}(t) = 2\max\left\{0,\, |\rho_{23}(t)|-\sqrt{\rho_{11}(t)\rho_{44}(t)}\right\} ,\label{3.14}
\end{equation}
which shows that the basic dynamical mechanism for entanglement creation in this system is the coherence $\rho_{23}(t)$. That is, the 
origin of the entanglement in the system can be traced back to the time evolution of the coherence $\rho_{23}(t)$. 

Utilizing the relation $\rho_{23}(t) = (u(t)-iv(t))/2$, the time evolution of the coherence $\rho_{23}(t)$ can be readily found from the 
solution of the Bloch equations (\ref{3.9}). The general solution for $u(t)$ and $v(t)$, valid for arbitrary initial conditions, 
is given by
\begin{eqnarray}
  u(t) &=& \bar{u}(t){\rm e}^{-\gamma t} \nonumber \\ 
  &=& \frac{{\rm e}^{-\gamma t}}{\alpha^{2}}\left[ 2\Omega_{12}{\cal A}
       + \delta_{12}\left(v_{0}\alpha\sin\alpha t +{\cal B}\cos\alpha t\right)\right] ,\nonumber \\
  v(t) &=& \bar{v}(t){\rm e}^{-\gamma t} = \frac{{\rm e}^{-\gamma t}}{\alpha}\left[v_{0}\alpha\cos\alpha t 
  +{\cal B}\sin\alpha t\right]  ,\label{3.15} 
\end{eqnarray}
where $w_{0}\equiv w(0)$, $u_{0}\equiv u(0)$ and $v_{0}\equiv v(0)$ determine the initial population inversion and coherences in the 
system 
\begin{equation}
  {\cal A} = 2\Omega_{12}u_{0}-\delta_{12}w_{0} ,\qquad {\cal B}= \delta_{12}u_{0}+2\Omega_{12}w_{0} ,\label{3.16}
\end{equation}
and $\alpha =\sqrt{4\Omega_{12}^{2}+\delta_{12}^{2}}$ is the detuned Rabi frequency. It is also worthwhile to find the time evolution 
of the population inversion
\begin{eqnarray}
  w(t) &=& \bar{w}(t){\rm e}^{-\gamma t} = \frac{{\rm e}^{-\gamma t}}{\alpha^{2}}\left\{ -\delta_{12}{\cal A}\right. \nonumber \\
       &&\left. +  2\Omega_{12}\left(v_{0}\alpha\sin\alpha t +{\cal B}\cos\alpha t\right)\right\}  
       ,\label{3.17} 
\end{eqnarray}
which will allow us to use the Bloch vector picture to gain insight into the problem of entanglement creation in the system.
For a single-quantum excitation, $\rho_{11}(t)=\rho_{44}(t)=0$, and then the Bloch vector component $w(t)$ determines the population 
distribution among the levels $\ket 2$ and $\ket 3$, and through the relation 
\begin{eqnarray}
  u^{2}(t) +v^{2}(t) +w^{2}(t) = {\rm e}^{-2\gamma t} ,\label{3.18} 
\end{eqnarray}
the component, in turn, determines the entanglement
\begin{equation}
  {\cal C}(t) = \max\left\{0,\,{\rm e}^{-\gamma t}\sqrt{1-\bar{w}^{2}(t)}\right\} .\label{3.19}
\end{equation}
This implies that the concurrence can be completely determined by observing changes in the populations of the system's energy levels. 
When the population is in the level $\ket 2$ or in the level $\ket 3$, $\bar{w}(t)=\pm 1$, and then ${\cal C}(t)=0$, whereas 
${\cal C}(t)$ achieves its optimum value ${\cal C}(t)=1$ when $\bar{w}(t)=0$. Therefore, we can interpret the entanglement as 
a consequence of a distribution of the population among the energy levels.

In the complete Bloch vector picture, the Bloch vector makes a constant angle $\theta =\tan^{-1}(-2\Omega_{12}/\delta_{12})$ 
with $\vec{\Omega}_{B}$, it rotates around it, tracing out a circle on the Bloch sphere. When the Bloch vector does {\it not} 
intersect the "north pole" $\vec{B}=(0,0,1)$ or the "south pole" $\vec{B}=(0,0,-1)$, the inversion $\bar{w}(t)\neq \pm 1$ and then 
an entanglement between the atoms occurs.

It is clear from Eqs.~(\ref{3.17}) and (\ref{3.19}) that the temporal evolution of the concurrence can be quite different depending 
on the initial conditions. However, the most interesting aspects of the solutions (\ref{3.17}) and (\ref{3.19}) relate to the modifications
of the time evolution of the system produced by the detuning $\delta_{12}$, because these features are not encountered at all under 
the equal coupling constants and appear never to have been investigated before. This motivates our study of the effect of a spatial 
location of the atoms on the atom-atom entanglement.

\section{Entanglement dynamics}

To illustrate the influence of the initial conditions and spatial location of the atoms on the time development of the entanglement, 
we will use the solutions (\ref{3.17}) and (\ref{3.19}) to calculate the concurrence ${\cal C}(t)$ for some distances $r_{12}$ and 
time $t$. For a fixed $r_{12}$, we examine the time evolution of ${\cal C}(t)$ for the case where the atoms are first prepared in some 
pure state that can be specified by the Bloch vector components $u_{0}, v_{0}$ and $w_{0}$ at time $t=0$. 
In particular, we consider three different sets of initial pure states of the system corresponding to single-quantum excitations. 
In the first, the atom 1 is assumed to reside in its lower level $\ket{g_{1}}$ and the atom 2 in its upper level $\ket{e_{2}}$, i.e.
the initial conditions for the Bloch vector components are $w_{0}=1,\, u_{0}=v_{0}=0$. In the second, the system is assumed prepared 
in a pure single-quantum superposition (entangled) state in which the two atoms oscillate in the opposite phase 
\begin{equation}
  \ket\Psi = \frac{1}{\sqrt{2}}\left( \ket{g_{1}}\ket{e_{2}} +i\ket{e_{1}}\ket{g_{2}}\right) .\label{4.1}
\end{equation}
In this case, $v_{0}=1,\, u_{0}=w_{0}=0$. In the third, the system is assumed prepared in a pure superposition state with
the atoms oscillating with the same phase 
\begin{equation}
  \ket\Psi = \frac{1}{\sqrt{2}}\left( \ket{g_{1}}\ket{e_{2}} +\ket{e_{1}}\ket{g_{2}}\right) .\label{4.2}
\end{equation}
For this state, the initial conditions for the Bloch vector components are $u_{0}=1,\, v_{0}=w_{0}=0$.

\subsection{Preparation with $w_{0}=1$, $u_{0}=v_{0}=0$}\label{A}

This limit corresponds to an initial condition in which the atom 1 is in the lower level and the atom 2 is in the upper level
at time $t=0$, i.e. in terms of the atomic density matrix elements $\rho_{22}(0)=1$ and $\rho_{33}(0)=\rho_{23}(0)=\rho_{32}(0)=0$. 
In this limit, the concurrence ${\cal C}(t)=0$ at $t=0$, and at later times is given~by
\begin{eqnarray}
  {\cal C}(t) = {\rm e}^{-\gamma t}\sqrt{1-\left[1 - 
  \frac{8\Omega_{12}^{2}}{\alpha^{2}}\sin^{2}\left(\frac{1}{2}\alpha t\right)\right]^{2}} .\label{4.3} 
\end{eqnarray}
It is seen that the qualitative features of the transient development of the entanglement depend on whether or not $\delta_{12}=0$.
The concurrence sinusoidally varies with frequency $\alpha/2$  whose the origin is in the dipole-dipole interaction and the presence 
of the detuning $\delta_{12}$. For the particular case of equal coupling constants 
when $\delta_{12}=0$, this result is reduced to that obtained previously in Refs.~\cite{zg00,xl05}.
The most obvious effect of having $\delta_{12}$, i.e. unequal shifts of the atomic resonances, is that the 
the amplitude of the oscillating term is always less than 2. As a consequence, the entanglement may appear on the time scale twice as 
long as for the case of equal coupling constants. It is easy to see from Eq.~(\ref{4.3}) that for $\delta_{12}=0$, the concurrence 
vanishes periodically at times
\begin{eqnarray}
    t_{0} = n\pi/\alpha ,\qquad n=0,1,2,\ldots   ,\label{4.4} 
\end{eqnarray}
whereas for $\delta_{12}\neq 0$, the concurrence behaves quite differently such that it vanishes only at times
\begin{eqnarray}
    t_{\delta} = 2t_{0} = 2n\pi/\alpha ,\qquad n=0,1,2,\ldots   ,\label{4.5} 
\end{eqnarray}
i.e. the entanglement exists on the the time scale twice as long as for the case of equal coupling constants.
\begin{figure}[ht]
\includegraphics[height=5cm]{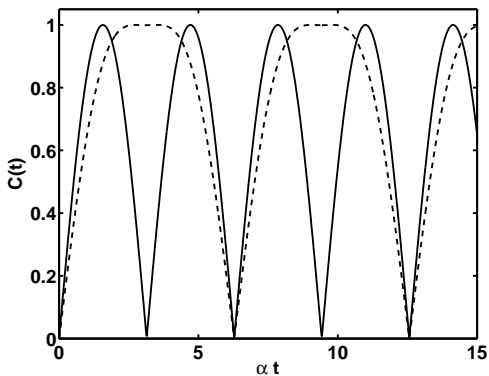}
\caption{Transient evolution of the concurrence ${\cal C}(t)$ for $\gamma =0$, atom $1$ located exactly at an antinode of the 
standing wave, and various locations $\delta r_{a}$ of the atom $2$ relative to an antinode of the standing wave:  
$\delta r_{a}=0$ (solid line), $\delta r_{a}= 0.18\lambda$ (dashed line). The system is initially in the state $\ket 2$.
The imperfect location $\delta r_{a}= 0.18\lambda$, leads to the detuning $\delta_{12} = 2\Omega_{12}$.}
\label{f1.1}
\end{figure}

The features described above are easily seen in Fig.~\ref{f1.1}, where we display the time evolution of ${\cal C}(t)$ for 
atom $1$ located exactly at an antinode of the standing wave and two different locations $\delta r_{a}$ of the atom $2$ relative to 
an antinode of the standing wave. According to Eq.~(\ref{2.11}), an imperfect or "nonideal" location of the atom $2$ inside the cavity 
mode leads to a nonzero detuning $\delta_{12}$. We assume a vanishing damping $\gamma =0$. This captures the essential 
dynamics of the system but makes no accounting of the dissipative process during the transient evolution. We see oscillations in the 
transient evolution of the entanglement that follow the Rabi flopping of population back and forth between the atoms. In other words, 
this oscillation reflects nutation of the atomic populations which, in turn, can be associated with the precession of the Bloch 
vector $\vec{B}$ about the driving field vector $\vec{\Omega}_{B}$ with frequency~$\alpha$. 

The most interesting feature of the transient entanglement seen in Fig.~\ref{f1.1} is that in the case of unequal coupling strengths, 
the initially unentangled system evolves into an entangled state, and remains in this state on the time scale twice as 
long as for the case of equal coupling strengths.
This rather surprising result can be understood in terms of spatial localization of the energy induced in the field by the initially 
excited atom. For equal coupling strengths the energy levels of the atoms are equally 
shifted by the amount $\delta_{1}=\delta_{2}$ due to the interaction with the cavity mode. In this case the induced energy by the 
first atom oscillates with frequency $\alpha$ such that at the particular times $t_{n}=n\pi/\alpha \ (n=0,1,2,\ldots )$ is fully 
absorbed by the second atom. Since at these times the energy is well localized in space as being completely absorbed 
by the localized atoms, the entanglement, which results from an unlocalized energy, is zero. The situation changes when 
$g_{1}\neq g_{2}$. According to (\ref{3.15}), in this case the energy levels of the atoms are unequally shifted 
by the interaction with the cavity mode. Due to the frequency mismatch, the energy induced by the atom~$1$ is not fully absorbed 
by the atom $2$, leading to a partial spatial delocalization of the photon at discrete times $t=n\pi/\alpha$, 
where $n=1,3,5,\ldots$. Consequently, at these times a partial entanglement is observed. The entanglement vanishes every time the 
excitation returns to its initial state, i.e. when it returns to atom~$2$.

\subsection{Preparation with $v_{0}=1$, $u_{0}=w_{0}=0$}

In this limit the initial population distributes equally between the levels $\ket 2$ and $\ket 3$, i.e. in terms of the bare states 
density matrix elements are $\rho_{22}(0)=\rho_{33}(0)=\rho_{23}(0)=\rho_{32}(0)=1/2$. It then follows from Eqs.~(\ref{3.14}) 
and~(\ref{3.15}) that the concurrence ${\cal C}(t)=1$ at $t=0$, and its time evolution is
\begin{eqnarray}
  {\cal C}(t) = {\rm e}^{-\gamma t}\sqrt{1-\frac{4\Omega_{12}^{2}}{\alpha^{2}} \sin^{2}\alpha t} .\label{4.6} 
\end{eqnarray}
We note immediately that the amplitude of the oscillating term that is equal to the population inversion is always less than unity 
provided $\delta_{12}\neq 0$. Thus, complete transfer of the population between the atoms cannot be achieved. In terms of the atomic 
excitation, none of the atoms can be completely inverted. This means that a part of the initial energy is always delocalized. 
Consequently, the system initially 
prepared in the entangled state of the form (\ref{4.1}) will remain entangled for all times.
We emphasize 
that this feature arises from the presence of a non-zero detuning $\delta_{12}$. That is the reason why the continuous in time entanglement 
is observed. Thus, this system may produce continuous in time atom-atom entanglement through having different values of the coupling 
constants $g_{1}$ and $g_{2}$.
\begin{figure}[ht]
\includegraphics[height=5cm]{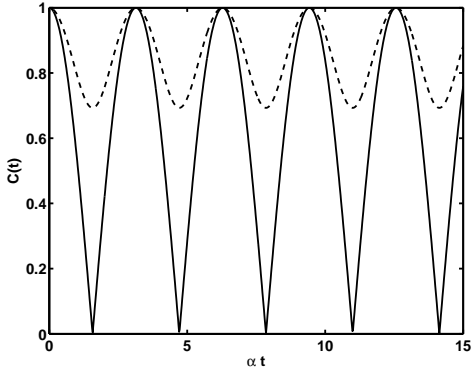}
\caption{Concurrence ${\cal C}(t)$ as a function of time for $\gamma =0$, atom $1$ located exactly at an antinode of the 
standing wave, and  different $\delta r_{a}$:  
$\delta r_{a}=0$ (solid line), $\delta r_{a}= 0.18\lambda$ (dashed line). The system is initially in the state 
$v_{0}=1$, $u_{0}=w_{0}=0$.}
\label{f1.2}
\end{figure} 

This behavior is illustrated in  Fig.~\ref{f1.2}, where ${\cal C}(t)$ is plotted for various values of $\delta_{12}$ and the 
initial superposition state~(\ref{4.1}). For $t=0$ the system is maximally entangled due to our choice of the initial state. Immediately
afterwards, the entanglement begins to decrease because of the coherent oscillation of the atomic excitation. For the case $\delta_{12}=0$,
the system becomes unentangled periodically at the times $t=n\pi/2$, whereas for $\delta_{12}\neq 0$, the periodic minima (nodes) of 
${\cal C}(t)$ are reduced in magnitude resulting in a nonzero entanglement present for all times.

The above analysis show that, rather surprisingly, imperfect coupling of the atoms to the cavity mode may actually help generate 
continuous in time atom-atom entanglement, through unequal shifting of the atomic resonance frequencies.

\subsection{Preparation with $u_{0}=1$, $v_{0}=w_{0}=0$}

If the system is initially prepared in the superposition state of the form (\ref{4.2}), the concurrence ${\cal C}(t)=1$ at $t=0$, and
the time evolution of the concurrence found from Eqs.~(\ref{3.17}) and~(\ref{3.19}) is of the form
\begin{eqnarray}
  {\cal C}(t) = {\rm e}^{-\gamma t}\sqrt{1-\frac{16\Omega_{12}^{2}\delta_{12}^{2}}{\alpha^{4}}\sin^{4}\left(\frac{1}{2}\alpha t\right)}
   .\label{4.7} 
\end{eqnarray}
It follows from Eq.~(\ref{4.7}) that in the absence of $\delta_{12}$, i.e. when the atoms are in equivalent positions inside the cavity 
mode, the entanglement oscillation is completely suppressed. When $\delta_{12}\neq 0$ the concurrence varies periodically in time. 
The amplitude of the oscillating term that is equal to the population inversion is less than unity, unless $\delta_{12}=2\Omega_{12}$ 
and then the entanglement is completely quenched at the times $t = n\pi/\alpha$.
\begin{figure}[ht]
\includegraphics[height=5cm]{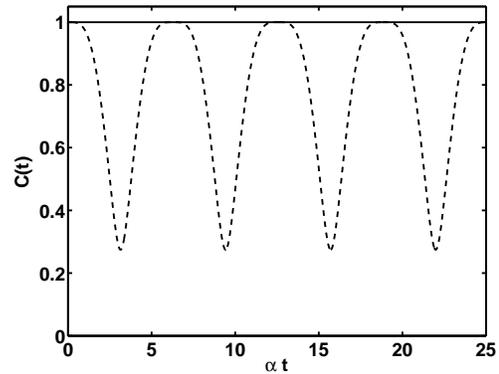}
\caption{Transient evolution of the concurrence ${\cal C}(t)$ for $\gamma =0$, atom $1$ located exactly at an antinode of the 
standing wave, and different $\delta r_{a}$: $\delta r_{a}=0$ (solid line), $\delta r_{a}= 0.17\lambda$ (dashed line). The system 
is initially in the state $u_{0}=1$, $v_{0}=w_{0}=0$.}
\label{f1.3}
\end{figure}

Figure~\ref{f1.3} shows the influence of the detuning $\delta_{12}$ on the time evolution of ${\cal C}(t)$. It is evident that 
$\delta_{12}$ significantly modifies the time evolution of ${\cal C}(t)$. For $\delta_{12}=0$ and vanishing damping, the entanglement is 
constant in time. Otherwise, when $\delta_{12}\neq 0$, the entanglement oscillates periodically and achieves the optimum value, 
${\cal C}(t)=1$ only at the particular times $t = 2n\pi/\alpha \ (n=0,1,2,\ldots)$. It can also vanish at the times $t = n\pi/\alpha$.
However, this happens only for a particular separation $r_{12}$ for which $\delta_{12}=2\Omega_{12}$. Thus, for positions of the atoms 
for which $\delta_{12}\neq 2\Omega_{12}$, entanglement is seen to occur over all times.

This behavior of the entanglement is linked to the fact that with the pure entangled state preparation of the form (\ref{4.2}) the Bloch 
vector and the dipole-dipole field vector are initially parallel. For $\delta_{12}=0$, the Bloch vector $\vec{B}$ is effectively locked 
to the field vector, i.e. $\vec{B}\times\vec{\Omega}_{B} =0$, and does not precess as it evolves on a time scale given by the spontaneous 
emission rate. In analogy to the spin locking effect, we may call this phenomenon as entanglement locking.
When $\delta_{12}\neq 0$, the concurrence oscillates in time at frequencies $\alpha$ and $\alpha/2$. In the Bloch picture this 
corresponds to the fact that for $t>0$ the Bloch vector $\vec{B}$ is no longer aligned along the $\vec{\Omega}_{B}$ vector$-$precession
of the Bloch vector translates into an oscillatory entanglement. 

Finally, let us examine in greater details the case $\delta_{12}=2\Omega_{12}$. We have seen that under this specific condition 
the concurrence exhibits interesting features. For example, in the case A, this is the value of $\delta_{12}$ at which
the entanglement achieves the optimum value with imperfect matching. In the case B, the entanglement is 
always greater than $50$\%, and in the case C, the population is inverted for all times, and consequently the entanglement can be completely 
quenched at some discrete times.
The reason is that at this value of $\delta_{12}$, the field vector $\vec{\Omega}_{B}$ is in the $uw$ plane, angle $\theta =-\pi/4$ from
the north pole $(w=1)$ of the Bloch sphere. The Bloch vector processes about a cone whose opening angle $\theta$ depends on the initial
conditions. In the case A, the the Bloch vector makes a constant angle $\theta =\pi/4$ with $\vec{\Omega}_{B}$, whereas in the
case C, it makes a constant angle $\theta =\pi/4$ with $-\vec{\Omega}_{B}$. Thus, in these two cases, the Bloch vector rotates in a
quarter of the Bloch sphere such that it can reach one of the poles, $w=\pm 1$.
It therefore appears that in the case A, the Bloch vector when processing about $\vec{\Omega}_{B}$ it varies from $w=0$ to $w=1$,
i.e. regularly reaches the north pole, but in the case B, it varies from $w=0$ to $w=-1$, i.e. regularly reaches the south pole. 
The case C is different, the Bloch vector makes an angle $\theta =\pi/2$ with $\vec{\Omega}_{B}$, so that it 
processes about $\vec{\Omega}_{B}$ in such a way that it
rotates from $w=-1/\sqrt{2}$ to $w=1/\sqrt{2}$ such that it never achieve the poles $w=\pm 1$.

\section{Entanglement diffraction}

One of the more interesting aspects of the transient entanglement demonstrated in the previous section is its dependence on the
difference of the Stark shifts of the atomic transition frequencies. This difference arises from an imperfect coupling of the 
atoms to the cavity mode that is a consequence of nonequivalent positions of the atoms inside the cavity mode. We proceed here 
to present more detailed studies of the sensitivity of the transient entanglement to position of the atoms inside a standing-wave 
cavity mode. We choose the initial conditions to be $w_{0}=1, u_{0}=v_{0}=0$. When Eqs.~(\ref{2.5}) and (\ref{2.11}) are used 
in Eq.~(\ref{4.3}), we readily find that the variation of the concurrence with position of the atoms in the standing wave is given by
\begin{eqnarray}
  {\cal C}(r_{12}) &=& {\rm e}^{-\Gamma \tau}\left\{\left(\frac{\sin d}{d}\right)^{2}\tau^{2}\right. \nonumber \\
  &+&\left. \left(\frac{\sin \frac{1}{2}d}{\frac{1}{2}d}\right)^{4}\tau^{4}\sin^{4}kr_{12}
  \right\}^{\frac{1}{2}}|\cos kr_{12}| ,\label{5.1} 
\end{eqnarray}
where $d = (1+\cos^{2}kr_{12})\tau/2$ and we have introduced a scaled time variable
\begin{eqnarray}
    \tau =\frac{2g_{0}^{2}}{\Delta}t ,\label{5.2} 
\end{eqnarray}
and the dimensionless damping rate
\begin{eqnarray}
    \Gamma =\frac{\Delta}{2g_{0}^{2}}\gamma ,\label{5.3} 
\end{eqnarray}
both measured in units of $2g_{0}^{2}/\Delta$ which is always assumed to be much smaller than unity.

The concurrence (\ref{5.1}) exhibits an interesting modulation of the amplitude of the harmonic oscillation. One could naively think 
that a variation of the concurrence with $r_{12}$ should reveal the cosine form of the cavity mode function. However, Eq.~(\ref{5.1}) 
shows that the concurrence is not a simple cosine function of $r_{12}$. It is given by the product of two terms, one the absolute value 
of the cavity mode function $|\cos kr_{12}|$ and the other the time- and position-dependent diffraction structure. 
That is, the amplitude of the standing wave cavity mode is in the form 
of position and time dependent diffraction pattern. Only at very early times $(\tau\ll 1)$, the oscillations are not modulated by the 
diffraction pattern and the concurrence reduces to $|\cos kr_{12}|$, but for longer times, ${\cal C}(r_{12})$ may vary slower or faster 
than the cosine functions. Within the diffraction structure itself, the magnitude of the concurrence exhibits a succession of modes and
of antinodes. As a consequence, the entanglement may be completely quenched even for locations of the atom $2$ close to an 
antinode of the cavity mode, and alternatively may achieve its optimum value even for locations of the atom $2$ 
close to a node of the cavity mode. It is easy to show that ${\cal C}(r_{12})$ vanishes periodically whenever 
$\cos\left[\left(1+\cos^{2}kr_{12}\right)\tau/2\right] =1$, i.e. for discrete times
\begin{eqnarray}
    \tau = \frac{4n\pi}{1+\cos^{2}kr_{12}} ,\qquad n=1,2,3,\ldots \label{5.4} 
\end{eqnarray} 
We may establish a relation between the number of zeros in ${\cal C}(r_{12})$ and the time $\tau$ by an elementary argument.
Since $\cos kr_{12}$ varies between zero and one, the shortest time at which ${\cal C}(r_{12})$ achieves at least one zero is $\tau=2\pi$. 
Thus, there are no zeros in ${\cal C}(r_{12})$ for locations of the atom between the successive nodes of the cavity mode function, 
if $\tau <2\pi$. For a given $\tau$ the number of zeros is limited by the fact that $\cos^{2}kr_{12}\leq 1$. This imposes lower and
upper limit on $n$ in Eq.~(\ref{5.4}). The upper limit is important, since it determines the width of the region about antinodes
of the cavity function where the optimum entanglement occurs. It follows from Eq.~(\ref{5.4}) that the criterion for vanishing
entanglement is satisfied for
\begin{eqnarray}
    n \leq \frac{\tau}{2\pi} .\label{5.5} 
\end{eqnarray}
The largest $n$ satisfying this inequality corresponds to the largest value of $\cos^{2} kr_{12}$, and therefore
determines a node that is the closest to the antinode of the cavity mode function, so that it
determines the width of the main peak of the diffraction pattern.

\begin{figure}[ht]
\includegraphics[height=5.5cm]{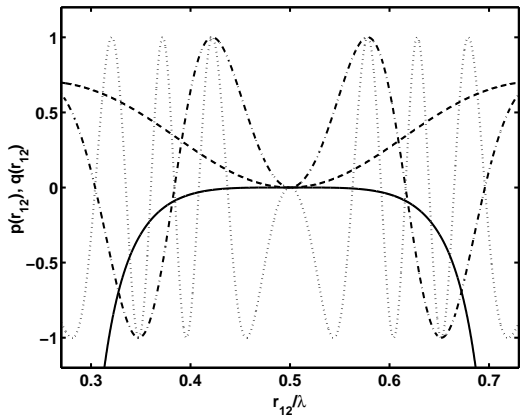}
\caption{The parameters $q(r_{12})$ (solid line) and $p(r_{12})$ plotted as a function of $r_{12}$ for different times $\tau$:
$\tau_{1}=\pi/2$ (dashed line), (b) $\tau_{5}=9\pi/2$ (dashed-dotted line), (c) $\tau_{10}=27\pi/2$ (dotted line). The choice of 
the particular values of $\tau$ corresponds to optimum entanglement observed for the idealized case of $g_{1}=g_{2}$.}
\label{f1.4}
\end{figure}

More interesting is a possibility of obtaining the optimum entanglement when the atom $2$ is located between a node and a 
successive antinode of the cavity mode. The optimum entanglement occurs at the locations of the most intense maxima of the concurrence.
However, the locations of the maxima are not given by a simple relation.
For a given $\tau$, the concurrence (\ref{5.1}) achieves the optimum value ${\cal C}(r_{12})=1$ whenever 
\begin{eqnarray}
    \cos\left[\frac{1}{2}\left(1+\cos^{2}kr_{12}\right)\tau \right] 
    = -\left(\frac{\sin^{2}kr_{12}}{2\cos kr_{12}}\right)^{2} .\label{5.6} 
\end{eqnarray}
This is not a simple relation, and we solve this equation graphically as follows. Introducing the notation 
\begin{eqnarray}
    p(r_{12}) &=& \cos\left[\frac{1}{2}\left(1+\cos^{2}kr_{12}\right)\tau\right] ,\nonumber \\
    q(r_{12}) &=& -\left(\frac{\sin^{2}kr_{12}}{2\cos kr_{12}}\right)^{2} ,\label{5.7} 
\end{eqnarray}
we find solutions of the equation $p(r_{12})=q(r_{12})$ by plotting separately $p(r_{12})$ and $q(r_{12})$. The functions 
$p(r_{12})$ and $q(r_{12})$ are shown in Fig.~\ref{f1.4}.
The intersection points of the two curves give the solutions of the equation (\ref{5.6}). At these points the system attains the optimum
entanglement. We see from the figure that the 
equation (\ref{5.6}) is satisfied only for discrete values of $r_{12}$. The number of solutions, which gives us the number of the 
optimum that the concurrence may achieve, depends on time $\tau$. Rather than examine the situation at all times we will look only 
at the particular times $\tau_{n} = n\pi/2$, corresponding to the evolution intervals the optimal entanglement is obtained for the idealized 
case of $g_{1}=g_{2}$. For $\tau =\pi/2$ there is only one solution corresponding to the position of the atom $2$ precisely 
at the antinode of the cavity mode function. For a longer time $\tau =9\pi/2$ there are three solutions and the number of solutions 
increases with $\tau$.

\begin{figure}[ht]
\includegraphics[height=6.5cm]{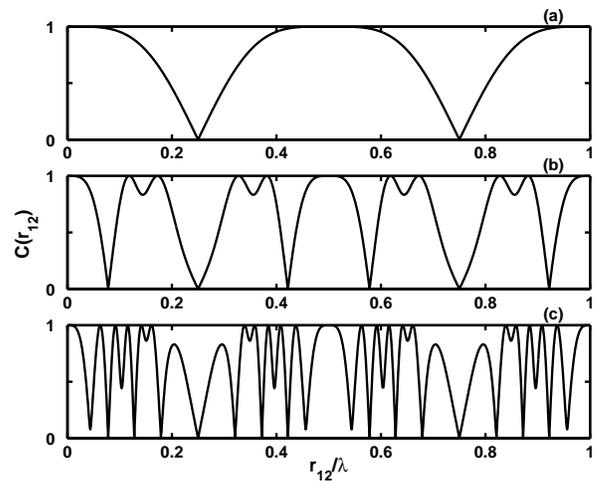}
\caption{Concurrence ${\cal C}(r_{12})$ as a function of the separation between the atoms when $\gamma =0$, the system is initially 
in the lower level, $w_{0}=1, u_{0}=v_{0}=0$, and (a) $\tau =\pi/2$, (b) $\tau =9\pi/2$, (c) $\tau =27\pi/2$.}
\label{f1.5}
\end{figure}

Figure~\ref{f1.5} shows ${\cal C}(t_{n})$ as a function of $r_{12}/\lambda$ for different times $\tau$. 
For a short time the entanglement is seen to occur over a wide range of positions centered about the antinodes of the cavity mode.
The concurrence is a bell-shaped function of position without any oscillation. As time progresses, oscillations appear and consequently 
the region of $r_{12}$ where the optimum entanglement occurs, becomes narrower. The evolution of ${\cal C}(r_{12})$ tends to become 
increasingly oscillatory with $r_{12}$ as time increases, and the optimum entanglement occurs in a still more restricted range of $r_{12}$.
As a result, the atom-atom entanglement oscillates with position faster than the cosine function, and the oscillations are more dramatic 
for larger times.
\begin{figure}[ht]
\includegraphics[height=6.5cm]{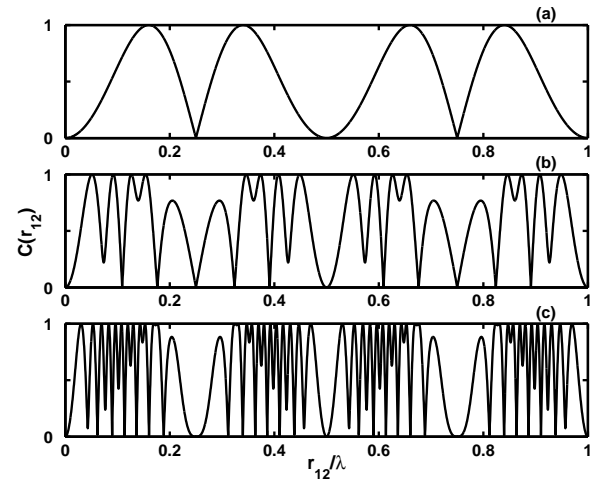}
\caption{Concurrence ${\cal C}(r_{12})$ as a function of the separation between the atoms when $\gamma =0$, the system is initially 
in the lower level, $w_{0}=1, u_{0}=v_{0}=0$, and (a) $\tau =\pi$, (b) $\tau =10\pi$, (c) $\tau =28\pi$.}
\label{f1.6}
\end{figure}

In Fig.~\ref{f1.6}, we plot ${\cal C}(t_{n})$ as a function of $r_{12}/\lambda$ for three values of time $\tau$ corresponding to 
the evolution intervals at which the entanglement is quenched for the idealized case of $g_{1}=g_{2}$. Here we observe, that the 
concurrence actually evolves with the position of the atom leading to the appearance of what we may call an inverse diffraction 
pattern. Note that the concurrence vanishes for locations of the atom precisely at the antinodes of the cavity mode, and may achieve 
its optimum value at locations of the atom close to the nodes of the cavity mode.

This effect can be understood as a consequence of the uncertainty relation between the evolution time and energy~\cite{fs05,kek06}. 
For the increasing time the uncertainty of the energy decreases which means that the energy becomes more localized. The increase in the 
localization of the energy results in a degradation of the entanglement. In other words, with increasing time, one can 
in principle obtain more information about the localization of the atoms inside the cavity mode.

\section{Summary}

We have investigated the process involved in the entanglement creation between two distant atoms coupled to a single-mode cavity field.
Unlike previous publications, we have included a possible variation of the coupling constant $g(\vec{r})$ with the location of the atoms 
in a standing-wave cavity mode. We have found that the entanglement creation in a complex two-qubit system can be modeled in terms of 
the coherent dynamics of a simple single-qubit system driven by a coherent field. Effectively, we have shown how to engineer two coupled 
qubits whose the dynamics are analogous to that of a driven single two-level system. We have obtained analytical expressions for the 
concurrence and have shown some new properties of the entanglement that are not met in the idealized case of equal coupling constants
appear for unequal or "imperfect" coupling constants. In particular, the degree of entanglement and the time it takes for the concurrence 
to reach its optimum value is a sensitive function of the position of the atoms inside the cavity mode. Characterizing the system by the 
Bloch vector components, we have examined the parameter ranges in which entanglement can take place for all times. 
We have demonstrated that a spatial variation of the coupling constant affects localization of the energy induced in the field by 
interacting atoms that leads to a long-lived entanglement. The consequence of this imperfection is that under certain initial conditions, 
an initially entangled system may remain entanglement for all times.

Finally, we have shown that the variation of the concurrence with the position of the atoms is that of the cavity mode function multiplied
by a time-dependent diffraction pattern. The diffraction formula shows explicitly the trend of the modification of the entanglement with
the localization of the atoms when the observation time increases. For a short time the entanglement is seen to occur over a wide range 
of positions centered about the antinodes of the cavity mode. As time progresses, oscillations appear and consequently the spatial region
where the optimum entanglement occurs, becomes narrower. This effect has been explained in terms of the quantum property of complementarity, 
which is manifested as a tradeoff between the knowledge of energy of the exchanged photon versus the evolution time of the~system.\\

\section*{ACKNOWLEDGMENTS}

This work was supported by the Australian Research Council.

\end{document}